\documentstyle[epsf]{mn}

\newif\ifAMStwofonts

\def\simlt{\lower.5ex\hbox{$\; \buildrel < \over \sim \;$}}
\def\simgt{\lower.5ex\hbox{$\; \buildrel > \over \sim \;$}}

\newcommand {\R}{{\mathcal R}}
\newcommand {\B}{{\mathcal B}}
\newcommand {\Like}{{\mathcal L}}
\newcommand {\zdec}{z_{dec}}
\newcommand {\ee}{\end{equation}}

\ifoldfss
  \ifCUPmtlplainloaded \else
    \NewTextAlphabet{textbfit} {cmbxti10} {}
    \NewTextAlphabet{textbfss} {cmssbx10} {}
    \NewMathAlphabet{mathbfit} {cmbxti10} {} 
    \NewMathAlphabet{mathbfss} {cmssbx10} {} 
  \fi
  \ifAMStwofonts
    \ifCUPmtlplainloaded \else
      \NewSymbolFont{upmath} {eurm10}
      \NewSymbolFont{AMSa} {msam10}
      \NewMathSymbol{\upi}     {0}{upmath}{19}
      \NewMathSymbol{\umu}     {0}{upmath}{16}
      \NewMathSymbol{\upartial}{0}{upmath}{40}
      \NewMathSymbol{\leqslant}{3}{AMSa}{36}
      \NewMathSymbol{\geqslant}{3}{AMSa}{3E}

       \let\le=\leqslant
       
    \fi
  \fi
\fi 

\ifnfssone
  \newmathalphabet{\mathit}
  \addtoversion{normal}{\mathit}{cmr}{m}{it}
  \addtoversion{bold}{\mathit}{cmr}{bx}{it}
  \newmathalphabet{\mathbfit} 
  \addtoversion{normal}{\mathbfit}{cmr}{bx}{it}
  \addtoversion{bold}{\mathbfit}{cmr}{bx}{it}
  \newmathalphabet{\mathbfss} 
  \addtoversion{normal}{\mathbfss}{cmss}{bx}{n}
  \addtoversion{bold}{\mathbfss}{cmss}{bx}{n}
  \ifAMStwofonts
    \ifCUPmtlplainloaded \else
      \UseAMStwoboldmath
      \makeatletter
      \new@mathgroup\upmath@group
      \define@mathgroup\mv@normal\upmath@group{eur}{m}{n}
      \define@mathgroup\mv@bold\upmath@group{eur}{b}{n}
      \edef\UPM{\hexnumber\upmath@group}
      \new@mathgroup\amsa@group
      \define@mathgroup\mv@normal\amsa@group{msa}{m}{n}
      \define@mathgroup\mv@bold\amsa@group{msa}{m}{n}
      \edef\AMSa{\hexnumber\amsa@group}
      \makeatother
      \mathchardef\upi="0\UPM19
      \mathchardef\umu="0\UPM16
      \mathchardef\upartial="0\UPM40
      \mathchardef\leqslant="3\AMSa36
      \mathchardef\geqslant="3\AMSa3E

       \let\le=\leqslant

    \fi
  \fi
\fi 

\ifnfsstwo
  \DeclareMathAlphabet{\mathbfit}{OT1}{cmr}{bx}{it}
  \SetMathAlphabet\mathbfit{bold}{OT1}{cmr}{bx}{it}
  \DeclareMathAlphabet{\mathbfss}{OT1}{cmss}{bx}{n}
  \SetMathAlphabet\mathbfss{bold}{OT1}{cmss}{bx}{n}
  \ifAMStwofonts
    \ifCUPmtlplainloaded \else
      \DeclareSymbolFont{UPM}{U}{eur}{m}{n}
      \SetSymbolFont{UPM}{bold}{U}{eur}{b}{n}
      \DeclareSymbolFont{AMSa}{U}{msa}{m}{n}
      \DeclareMathSymbol{\upi}{0}{UPM}{"19}
      \DeclareMathSymbol{\umu}{0}{UPM}{"16}
      \DeclareMathSymbol{\upartial}{0}{UPM}{"40}
      \DeclareMathSymbol{\leqslant}{3}{AMSa}{"36}
      \DeclareMathSymbol{\geqslant}{3}{AMSa}{"3E}

       \let\le=\leqslant

    \fi
  \fi
\fi 

\ifCUPmtlplainloaded \else
  \ifAMStwofonts \else 
    \def\upi{\pi}
    \def\umu{\mu}
    \def\upartial{\partial}
  \fi
\fi

\title[The impact of extra relativistic particles on CMB]
	{The Impact of an Extra Background of Relativistic Particles on
the Cosmological Parameters derived from Microwave Background Anisotropies}
\author[Bowen et al.]
{Rebecca Bowen$^1$, Steen H. Hansen$^1$, Alessandro Melchiorri$^1$, Joseph Silk$^1$, \cr
Roberto Trotta$^2$\\
$^1$ Nuclear and Astrophysics Laboratory, University of Oxford, 
Keble Road, Oxford, OX 3RH, UK\\
$^2$ D\'epartement de Physique Th\'eorique, Universit\'e de
  Gen\`eve, 24 quai Ernest Ansermet, CH-1211 Gen\`eve 4, Switzerland}
\pagerange{\pageref{firstpage}--\pageref{lastpage}}
\pubyear{2001}

\begin{document}

\maketitle

\label{firstpage}

\begin{abstract}
Recent estimates of cosmological parameters derived from Cosmic Microwave
Background (CMB) anisotropies are based on the assumption that we know
 the precise amount of energy density in relativistic particles in the
universe, $\omega_{rel}$, at all times.  There are, however, many
possible mechanisms that can undermine this assumption. In this paper
we investigate the effect that removing this assumption has on
the determination of the various cosmological parameters.  We obtain
fairly general bounds on the redshift of equality, $z_{eq}=
\omega_{m}/\omega_{rel}=3100_{-400}^{+600}$.  
We show that $\omega_{rel}$ is nearly degenerate with the 
amount of energy in matter, $\omega_m$, and that
its inclusion in CMB parameter estimation also affects the present
constraints on other parameters such as the curvature or the scalar
spectral index of primordial fluctuations. This degeneracy 
has the effect of limiting the precision of parameter estimation from the
MAP satellite, but it can be broken by measurements on smaller
scales such as those provided by the Planck satellite mission.
\end{abstract}


\section{Introduction}
Our knowledge of the cosmological parameters has increased
dramatically with the release of recent cosmic microwave background
(CMB) observations (Netterfield et al. 2001; Lee et al. 2001;
Halverson et al. 2001).  Recent analyses of these datasets (de
Bernardis et al. 2001; Pryke et al. 2001; Wang, Tegmark \& Zaldarriaga 
2001) have
reported strong new  constraints on various parameters including the
curvature of the universe and the amount of baryonic and dark matter.
The precise determination from the CMB of other parameters such as the
cosmological constant or the spectral index of primordial fluctuations
can be limited by various degeneracies, and such degeneracies are best
lifted by combining CMB data with either supernova (SN) data
(Garnavich et al. 1998; Perlmutter et al. 1997) or large scale
structure (LSS) surveys, such as PSCz (Saunders et al. 2000), 2dF
(Percival et al. 2001) or Lyman-$\alpha$ (Croft et al. 2001) data.  
At present, the values obtained from CMB measurements under the
assumption of purely adiabatic fluctuations are consistent with 
the generic predictions of the inflationary scenario, 
$n_s\sim 1$ and $\Omega_{tot}=1$ (Linde 1990), and with the standard 
big-bang nucleosynthesis bound, 
$\Omega_bh^2=0.020\pm0.002$ (Burles et al. 2001; Esposito et al. 2001).
All of these observations  converge towards a consistent picture
of our universe, providing strong support for the inflationary scenario
as the mechanism which generated the initial
conditions for structure formation.  

The derivation of the cosmological parameters from CMB is, however, an
{\it indirect} measurement, relying on the assumptions of a
theoretical scenario.  For this reason, recent efforts have been
made to study the effects of the removal of some of these assumptions.
For example, a scale-invariant background of gravity waves, generally
expected to be small, has been included in the analysis of Kinney,
Melchiorri \& Riotto (2000), 
Wang et al. (2001) and Efstathiou et al. (2001), with
important consequences for parameter estimation.  A scale-dependence of
the spectral index has been included in the analysis of Griffiths, Silk
\& Zaroubi (2001), Santos et al. (2001) and Hannestad et al. (2001).
Furthermore, in Bucher, Moodley \& Turok (2000), Trotta, Riazuelo \& Durrer
(2001) and Amendola
et al. (2001), the effects of including isocurvature modes, which 
naturally arise in the most general inflationary scenarios, have been
studied,  with the finding that the inclusion of these modes can significantly
alter the CMB result. Even more drastic alterations have been proposed
in Bouchet et al. (2001) and Durrer, Kunz \& Melchiorri (2001).

All the above modifications primarily affect the constraints on the
curvature, on the physical baryon density parameter,
$\omega_b=\Omega_bh^2$, and the scalar spectral index $n_s$.  

In this paper we study another possible modification to the
standard scenario, namely variations in the parameter $\omega_{rel}$ which
describes the energy density of relativistic particles at times near
decoupling, $T \sim \mbox{0.1 eV}$.  CMB data analysis with variations in
this parameter has been recently undertaken by many authors (Hannestad
2000; Esposito et al. 2001; Orito et al. 2001; Hansen et al. 2001;
Kneller et al. 2001; Hannestad 2001; Zentner \& Walker 2001), 
giving rather crude upper bounds,
significantly improved only by including priors on the age of the
universe or by including supernovae (SN) or large scale structure (LSS) 
data.  It is worth emphasizing
that there is little difference in the bounds obtained on $N_{eff}$,
the effective number of relativistic species, 
between old and recent CMB data because of the degeneracy which we
will describe in detail below.  We  focus here on the effects that
the inclusion of this parameter, $\omega_{rel}$, has on the
constraints of the remaining parameters in the context of purely
adiabatic models.

As we will show below (and as observed previously, see e.g. 
Hu et al. (1999)) there is a strong degeneracy between
$\omega_{rel}$ and $\omega_m$. This is important, because
an accurate determination of $\omega_m$ from CMB observations (and of
$\Omega_m$ by including the Hubble Space Telescope result
$h=0.72\pm0.08$) can be useful for a large number of reasons.  First
of all, determining $\omega_{cdm}=\omega_m-\omega_b$ can shed new
light on the nature of dark matter. The thermally averaged
product of cross-section and  thermal  velocity of the dark matter candidate is related
to $\omega_m$, and this relation can be used to analyze the
implications for the mass spectra in versions of the Supersymmetric
Standard Model (see e.g. Barger \& Kao 2001, Djouadi, Drees \& Kneur 2001,
Ellis, Nanopoulos \& Olive 2001).  
The value of $\Omega_m$ can be determined in an
independent way from the mass-to-light ratios of clusters (Turner
2001), and the present value is $0.1 < \Omega_m < 0.2$ (Carlberg et
al. 1997; Bahcall et al. 2000).
Furthermore, a precise measurement of $\Omega_m$ will be a key
input for determining the redshift evolution of the equation of state
parameter $w(z)$ and thus discriminating between different quintessential 
scenarios (see e.g. Weller and Albrecht 2001). 

This paper is structured as follows.  In the next section, we
briefly review various physical mechanisms that can lead to a change
in $\omega_{rel}$ with respect to the standard value. In section 3, we
illustrate how the CMB angular power spectrum depends on this
parameter and identify possible degeneracies with other
parameters. In section 4, we present a likelihood analysis from
the most recent CMB data and show which of the present constraints on
the various parameters are affected by variations in
$\omega_{rel}$.  Section 5 forecasts the precision in the estimation
of cosmological parameters for the future
space missions MAP and Planck. Finally, in section 6, we discuss
our results and present our conclusions.

\section{Effective number of relativistic species}

In the standard model $\omega_{rel}$ includes photons and neutrinos,
and it can be expressed as
\begin{equation}
\omega_{rel} = \omega_{\gamma} + N_{eff} \cdot \omega_{\nu}
\end{equation}
where $\omega_{\gamma}$ is the energy density in photons and $
\omega_{\nu}$ is the energy density in one active neutrino.  Measuring
$\omega_{rel}$ thus gives a direct observation on the effective number
of neutrinos, $N_{eff}$. Naturally there are only 3 active neutrinos,
and $N_{eff}$ is simply a convenient parametrization for the extra
possible relativistic degrees of freedom
\begin{equation} 
N_{eff} = 3 + \Delta N_{CMB} \, .
\end{equation}
Thus $\omega_{rel}$ includes energy density from all the relativistic
particles: photons, neutrinos, and additional hypothetical
relativistic particles such as a light majoron or a sterile neutrino.
Such hypothetical particles are strongly constrained from standard big
bang nucleosynthesis (BBN), where the allowed extra relativistic
degrees of freedom typically are expressed through the effective
number of neutrinos, $N_{eff} = 3 + \Delta N_{BBN}$.  BBN bounds are
typically about $\Delta N_{BBN} < 0.2 - 1.0$ (Burles et al. 1999; Lisi,
Sarkar \& Villante 1999).

One should, however, be careful when comparing the effective number of
neutrino degrees of freedom at the times of BBN and CMBR, since they
may be related by different physics (Hansen et al. 2001). This is
because the energy density in relativistic species may change from the
time of BBN ($T \sim$ MeV) to the time of last rescattering ($T \sim$ eV). For
instance, if one of the active neutrinos has a mass in the range eV $<
m <$ MeV and decays into sterile particles such as other neutrinos,
majorons etc. with lifetime $t(\mbox{BBN}) < \tau < t(\mbox{CMBR})$,
then the effective number of neutrinos at CMBR would be substantially
different from the number at BBN (White, Gelmini \& Silk 1995).  Such massive
active neutrinos, however, do not look too natural any longer in view
of the recent experimental results on neutrino oscillations (Fogli et
al. 2001; Gonzalez-Garcia et al. 2001), showing that all active neutrinos are
likely to have masses smaller than  an eV. One could instead consider
sterile neutrinos mixed with active ones which could be produced in
the early universe by scattering, and subsequently decay.  The mixing
angle must then be large enough to thermalize the sterile neutrinos
(Langacker 1989), and this can be expressed through the sterile to
active neutrino number density ratio $n_s / n_\nu \approx 4 {\cdot}
10^4 \sin^2 2\theta \ (m/\mbox{keV}) (10.75/g^*)^{3/2}$ (Dolgov \&
Hansen 2001), where $\theta$ is the mixing angle, and $g^*$ counts the
relativistic degrees of freedom. With $n_s/n_\nu$ of order unity we
use the decay time, $\tau \approx 10^{20} (\mbox{keV}/m)^5 /\sin^2
2\theta$ sec, and one finds, $\tau \approx 10^{17} (keV/m)^4 \, yr$,
which is much longer than the age of the universe for $m \sim
\mbox{keV}$, so they would certainly not have decayed at
$t(\mbox{CMBR})$. A sterile neutrino with mass of a few MeV
would seem to have the right decay time, $\tau \sim 10^5$~yr, but this is
excluded by standard BBN considerations (Kolb et al. 1991; Dolgov,
Hansen \& Semikoz 1998). 

Even though the simplest models predict that the relativistic degrees
of freedom are the same at BBN and CMB times, one could imagine more
inventive models such as quintessence (Albrect \& Skordis 2000;
Skordis \& Albrect 2001) which effectively could change $\Delta N$
between BBN and CMB (Bean, Hansen \& Melchiorri 2001).  
Naturally $\Delta N$ can be
both positive and negative.  For BBN, $\Delta N$ can be negative if the
electron neutrinos have a non-zero chemical potential (Kang \&
Steigman 1992; Kneller et al. 2001), or more generally with a
non-equilibrium electron neutrino distribution function (see
e.g. Hansen \& Villante 2000). To give an explicit (but highly exotic)
example of a different number of relativistic degrees of freedom
between BBN and CMB, one could consider the following scenario.
Imagine another 2 sterile neutrinos, one of which is essentially
massless and has a mixing angle with any of the active neutrinos just
big enough to bring it into equilibrium in the early universe, and one
with a mass of $m_{\nu_s}=3$ MeV and decay time
$\tau_{\nu_s}=0.1$ sec, in the decay channel $\nu_s \rightarrow \nu_e
+ \phi$, with $\phi$ a light scalar. The resulting non-equilibrium
electron neutrinos happen to exactly cancel the effect of the massless
sterile state, and hence we have $\Delta N_{BBN} = 0$. However, for
CMB the picture is much simpler, and we have just the stable sterile
state and the majoron, hence $\Delta N_{CMB} = 1.57$.  For CMB, one can
imagine a negative $\Delta N$ from decaying particles, where the decay
products are photons or electron/positrons which essentially increases
the photon temperature relative to the neutrino temperature
(Kaplinghat \& Turner 2001). Such a scenario naturally also dilutes
the baryon density, and the agreement on $\omega_b$ from BBN and CMB
gives a bound on how negative $\Delta N_{CMB}$ can be.  Considering
all these possibilities, we will therefore not make the usual
assumption, $\Delta N_{BBN} = \Delta N_{CMB}$, but instead consider
$\Delta N_{CMB}$ as a completely free parameter in the
following analysis.

The standard model value for $N_{eff}$ with 3 active neutrinos is
$3.044$.  This small correction arises from the combination of two
effects arising around the temperature $T \sim \mbox{MeV}$. These
effects are the finite temperature QED correction to the energy
density of the electromagnetic plasma (Heckler 1994), which gives
$\Delta N = 0.01$ (Lopez \& Turner 1999; Lopez et al. 1999).  If there
are more relativistic species than active neutrinos, then this effect
will be correspondingly higher (Steigman 2001).  The other effect
comes from neutrinos sharing in the energy density of the
annihilating electrons (Dicus et al. 1982), which gives $\Delta N =
0.034$ (Dolgov, Hansen \& Semikoz 1997, 1999; Esposito et
al. 2000). Thus one finds $N_{eff} = 3.044$.  It still remains to
accurately calculate these two effects simultaneously.

\section{CMB theory and degeneracies}

The structure of the $C_{\ell}$ spectrum depends essentially
on 4 cosmological parameters
\begin{equation}
\omega_b \, \, \, , \, \, \, \omega_m \, \, \, , \, \, \, \omega_{rel}
\, \, \, \mbox{and} \, \, \, {\cal R}\, \, \, ,
\label{eq:4par}
\end{equation}
the physical baryonic density $\omega_b \equiv \Omega_b h^2$, the energy
density in matter $\omega_m \equiv (\Omega_{cdm}+\Omega_b)h^2$, the energy
density in radiation $\omega_{rel}$ and the `shift' parameter 
$\R \equiv \ell_{ref} / \ell$, which gives the position of the 
acoustic peaks with respect to a flat, $\Omega_\Lambda = 0$ reference model.
Here $h$ denotes the Hubble parameter today, $H_0 \equiv 100h$ 
km s$^{-1}$ Mpc$^{-1}$, and $\Omega_{\Lambda}$ is the density 
parameter due to a cosmological constant, $\Omega_{\Lambda} \equiv 
\Lambda/3H_0^2$. 
In previous analyses (Efstathiou \& Bond
1999; Melchiorri \& Griffiths 2000 and references therein),
the parameter $\omega_{rel}$ has
been kept fixed to the standard value, while here we will allow it to
vary.  It is therefore convenient to write $ \omega_{rel} = 4.13 \cdot
10^{-5} (1 + 0.135 \cdot \Delta N)$ (taking $T_{CMB} = 2.726$ K),  
where $\Delta N$ is the excess number
of relativistic species with respect to the standard model, $N_{eff} =
3 + \Delta N$. The shift parameter $\R$ depends on
$\Omega_m \equiv \Omega_{cdm} + \Omega_b$, on the curvature $\Omega_k$ 
and on $\Omega_{rel}=\omega_{rel}/h^2$ through

\begin{eqnarray}
\label{eq:def_r}
\R &=& 2 \left( 1 - \frac{1}{\sqrt{1 + z_{dec}}} \right)  \nonumber \\
   && \times  \frac{\sqrt{| \Omega_k| }}{ \Omega_m}
       \frac{1}{\chi(y)} 
    \left[ \sqrt{\Omega_{rel} + 
    \frac{\Omega_m}{1 + \zdec} } - \sqrt{\Omega_{rel}} \right]  ,
\end{eqnarray}
where
\begin{eqnarray}
y &=&  \sqrt{|\Omega_k|}\int_0^{z_{dec}} \, dz\\
    && {[\Omega_{rel} (1+z)^4 + 
\Omega_m(1+z)^3+\Omega_k(1+z)^2+\Omega_{\Lambda}]^{-1/2}}. \nonumber
\end{eqnarray}

The function $\chi(y)$ depends on the curvature of the universe and is
$y$, $\sin(y)$ or $\sinh(y)$ for flat, closed or open models,
respectively.  Eq. (\ref{eq:def_r}) generalizes the expression for
$\R$ given in Melchiorri \& Griffiths (2001) to the case of
non-constant $ \Omega_{rel}$.

\begin{figure}
{\centerline{\vbox{\epsfxsize=8.5cm\epsfbox{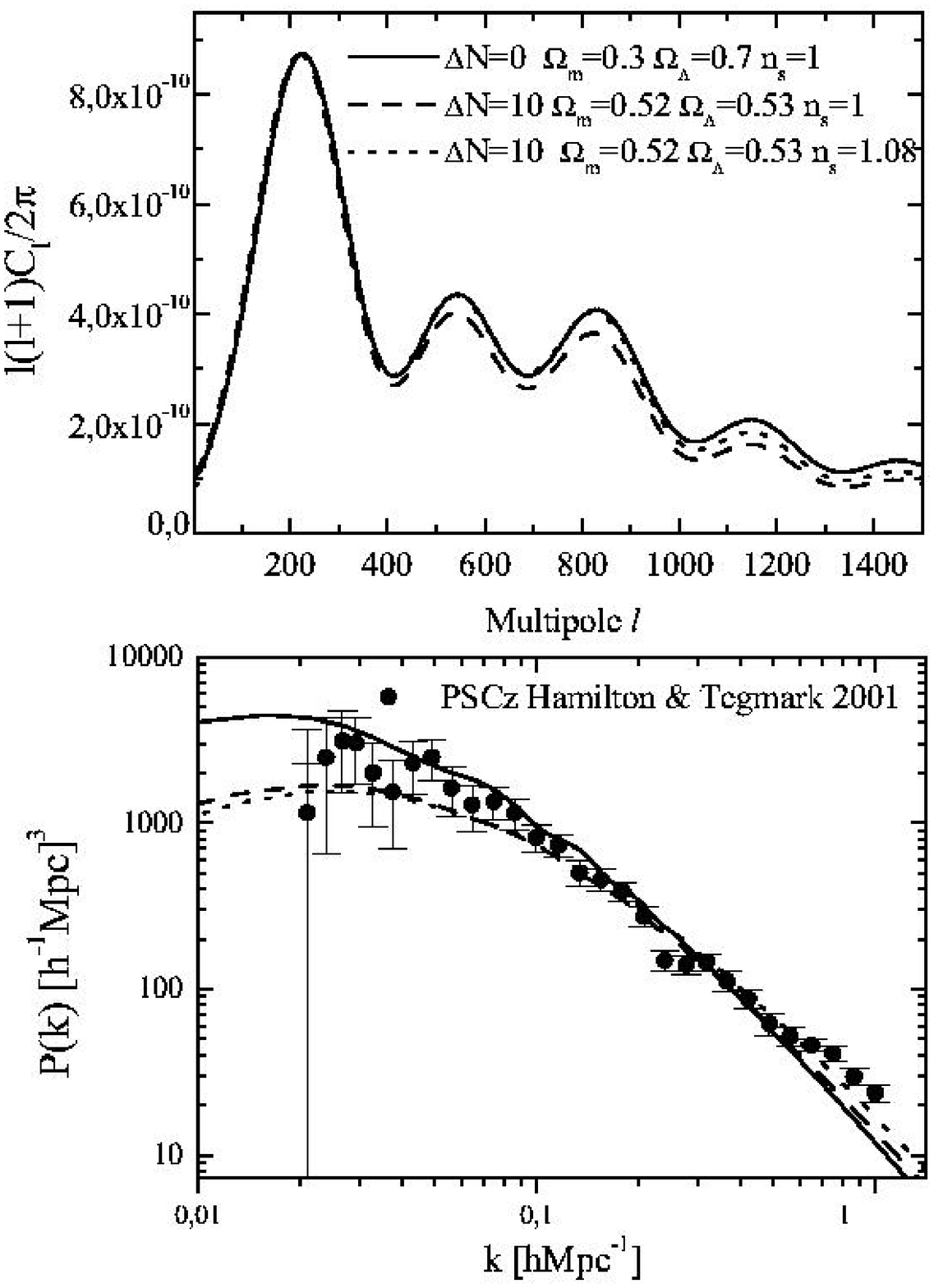}}} 
{\small
F{\scriptsize IG}.~1 --- Top panel: CMB 
degeneracies between cosmological models.
Keeping $z_{eq}, \omega_b$ and ${\cal R}$ fixed while varying 
$\Delta N$  produces nearly degenerate power spectra. The reference
model (black, solid) has $\Delta N = 0$, $\Omega_{tot} = 1.00$, 
$n_s = 1.00$; the nearly degenerate model (blue, dotted) has 
$\Delta N = 10$, $\Omega_{tot} = 1.05$, $n_s = 1.00$. The position
of the peaks is perfectly matched, only the relative height 
between the first and the other acoustic peaks is somewhat different
in this extreme example. The degeneracy can be further improved, at
least up to the third peak, by
raising the spectral index to $n_s = 1.08$ (red, dashed). 
Bottom panel: the matter power spectra of the models plotted in the
top panel together with the observed decorrelated power spectrum
from the PSCz survey (Hamilton and Tegmark 2000). The geometrical degeneracy
is now lifted.
\label{fig:degg}}
}
\end{figure}

By fixing the $4$ parameters given in (\ref{eq:4par}), or equivalently
the set $\omega_b$, the redshift of equality 
$z_{eq} \equiv \omega_m/\omega_{rel}$, 
$\Delta N$ and $\R$, one obtains a perfect
degeneracy for the CMB anisotropy power spectra on degree and
sub-degree angular scales. On larger angular scales, the 
degeneracy is broken by the late Integrated Sachs-Wolfe (ISW)
effect because of the different curvature and cosmological
constant content of the models.
From the practical point of view, however, it is still
very difficult to break the degeneracy, since measurements are
limited by ``cosmic variance'' on those scales, and because of
the possible contribution of gravitational waves.

Allowing $\Delta N$ to vary, but keeping constant the other
3 parameters $\omega_b$, $z_{eq}$, and $\R$, we obtain nearly
degenerate power spectra which we plot in Fig.\ 1, 
normalized to the first acoustic peak.
The degeneracy in the acoustic peaks region 
is now slightly spoiled by the
variation of the ratio $\Omega_{\gamma}/\Omega_{rel}$:
the different radiation content at decoupling induces
a larger (for $\Delta N > 0$) early ISW effect, which boosts 
the height of the first peak with respect to the other
acoustic peaks. Nevertheless, it is still impossible to 
distinguish between the different models 
with present CMB measurements and without external priors. 
Furthermore, a slight change in the scalar spectral index,
$n_s$, can reproduce a perfect degeneracy up to the third peak.

The main result of this is that, even with a measurement of the first
3 peaks in the angular spectrum, it is impossible to put bounds on 
$\omega_{rel}$ alone, even when fixing other parameters such as
$\omega_b$. Furthermore, since the degeneracy is mainly in $z_{eq}$,
the constraints on $\omega_m$ from CMB are also affected (see section
\ref{cmbanalysis}).

In Fig. 2 we plot the shift parameter $\R$
as a function of $\Delta N$, while fixing $\Omega_m=0.3$ and
$\Omega_{\Lambda}=0.7$. Increasing $\Delta N$ moves the peaks to smaller 
angular scales, even though the dependence of the shift parameter
on $\Delta N$ is rather mild. In order to compensate
this effect, one has to change the curvature by increasing
$\Omega_m$ and $\Omega_{\Lambda}$. We therefore conclude that
the present bounds on the curvature of the universe are weakly
affected by $\Delta N$. Nevertheless, when a positive (negative) 
$\Delta N$ is included in the analysis, the preferred models are 
shifted toward closed (open) universes.

\begin{figure}
{\centerline{\vbox{\epsfxsize=8.5cm\epsfbox{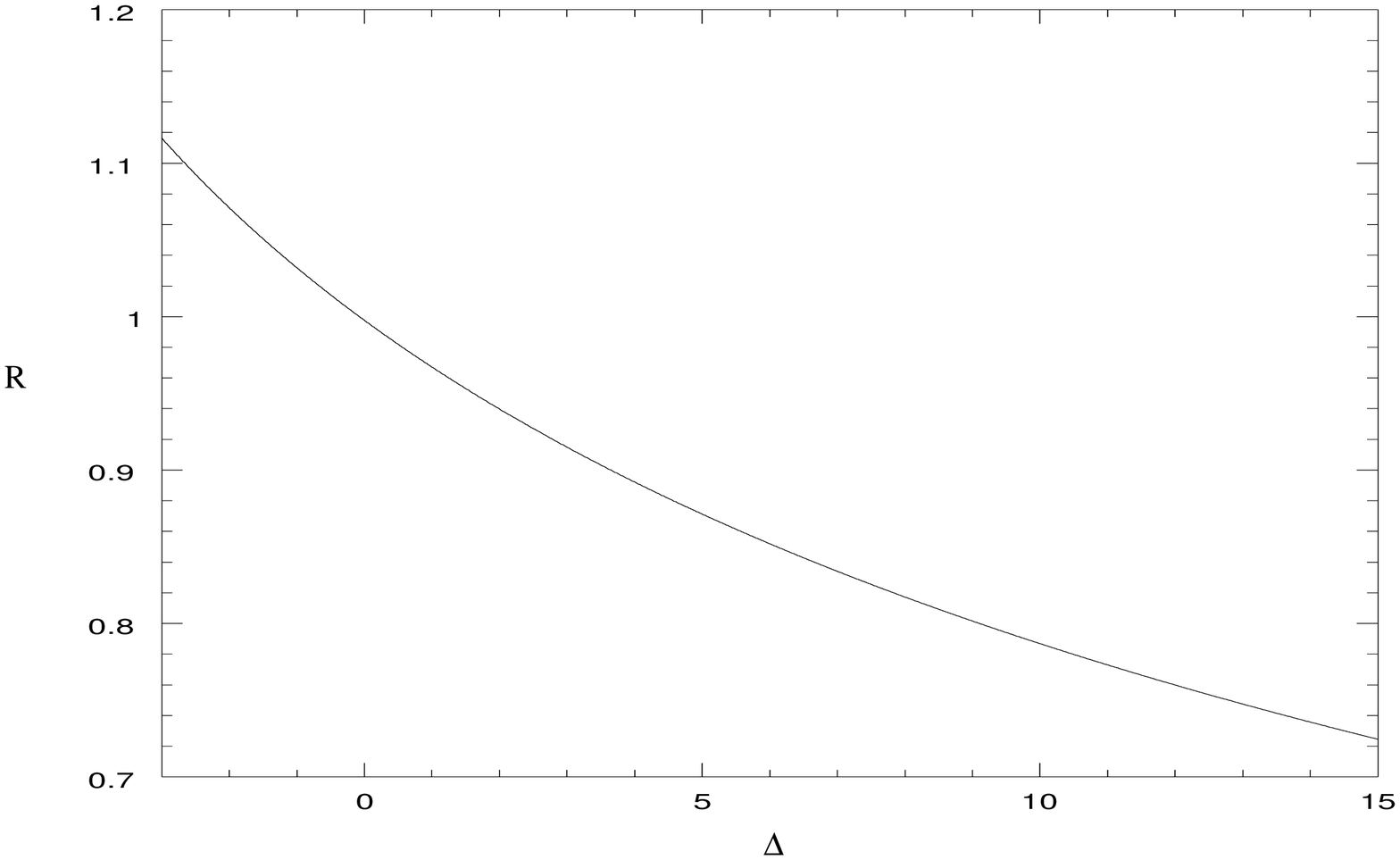}}}
{\small F{\scriptsize IG}.~2.--- 
${\cal R}$ as a function of $\Delta N$ with 
$\Omega_{\Lambda}=0.7$ and $\Omega_m=0.3$. The 
position of the peaks is only weakly affected by $\Delta N$.
\label{fig:R}}
}
\end{figure}

\section{CMB analysis}
\label{cmbanalysis}

In this section, we compare the recent
CMB observations with a set of models with cosmological parameters
sampled as follows: $0.1 < \Omega_{m} < 1.0$, $0.1<
\Omega_{rel}/\Omega_{rel}(\Delta N=0) < 3$, $0.015 < \Omega_{b} < 0.2$;
$0< \Omega_{\Lambda} < 1.0$ and $0.40 < h < 0.95$.  We vary the
spectral index of the primordial density perturbations within the
range $n_s=0.50, ..., 1.50$ and we re-scale the fluctuation amplitude
by a pre-factor $C_{10}$, in units of $C_{10}^{COBE}$.  We also
restrict our analysis to purely adiabatic, 
{\it flat} models ($\Omega_{tot}=1$) and we
add an external Gaussian prior on the Hubble parameter $h=0.65 \pm
0.2$.

The theoretical models are computed using the publicly available {\sc
cmbfast} program (Seljak \& Zaldarriaga 1996) and are compared with the recent
BOOMERanG-98, DASI and MAXIMA-1 results.  The power spectra from these
experiments were estimated in $19$, $9$ and $13$ bins respectively,
spanning the range $25 \le \ell \le 1100$.  We approximate the
experimental signal $C_B^{ex}$ inside the bin to be a Gaussian
variable, and we compute the corresponding theoretical value
$C_B^{th}$ by convolving the spectra computed by CMBFAST with the
respective window functions. When the window functions are not
available, as in the case of Boomerang-98, we use top-hat window
functions.  The likelihood for a given cosmological model is then
defined by $-2{\rm ln}
L=(C_B^{th}-C_B^{ex})M_{BB'}(C_{B'}^{th}-C_{B'}^{ex})$ where
$C_B^{th}$ ($C_B^{ex}$) is the theoretical (experimental) band power
and $M_{BB'}$ is the Gaussian curvature of the likelihood matrix at
the peak.  We consider $10 \%$, $4 \%$ and $4 \%$ Gaussian distributed
calibration errors (in $\mu$ K) for the BOOMERanG-98, DASI and MAXIMA-1
experiments respectively.  We also include the COBE data using Lloyd
Knox's RADPack packages.

In order to show the effect of the inclusion of $\omega_{rel}$ on the
estimation of the other parameters, we plot likelihood contours in the
$\omega_{rel}-\omega_{m}$, $\omega_{rel}-\omega_b$, $\omega_{rel}-n_s$
planes.

Proceeding as in Melchiorri et al. (2000), we calculate a likelihood
contour in those planes by finding the remaining 'nuisance' parameters
that maximize it. We then define our $68\%$, $95\%$ and $99\%$
confidence levels to be where the likelihood falls to $0.32$, $0.05$
and $0.01$ of its peak value, as would be the case for a 2-dimensional
Gaussian.

\begin{figure}
{\centerline{\vbox{\epsfxsize=6.0cm\epsfbox{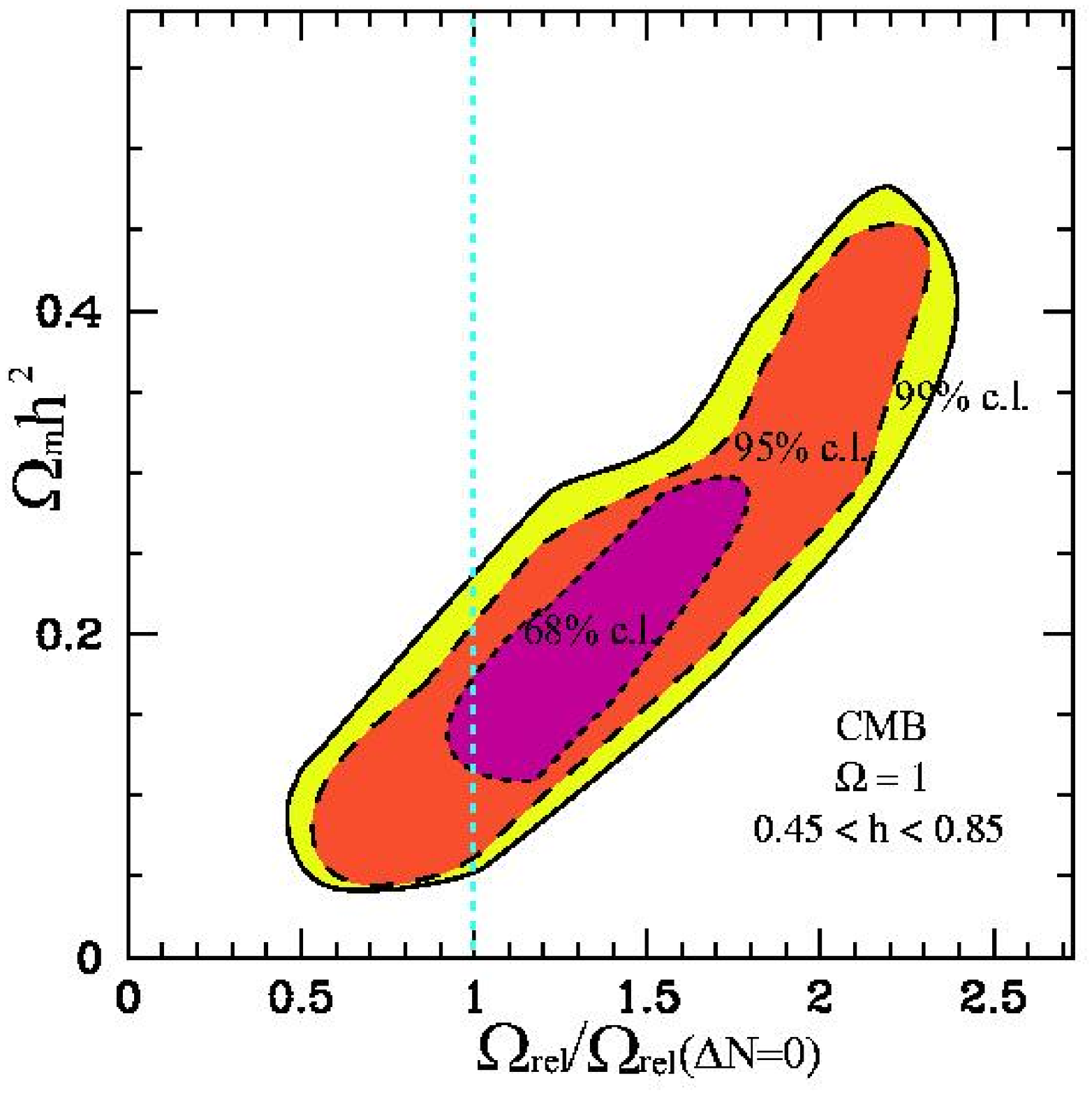}}}
{\centerline{\vbox{\epsfxsize=6.0cm\epsfbox{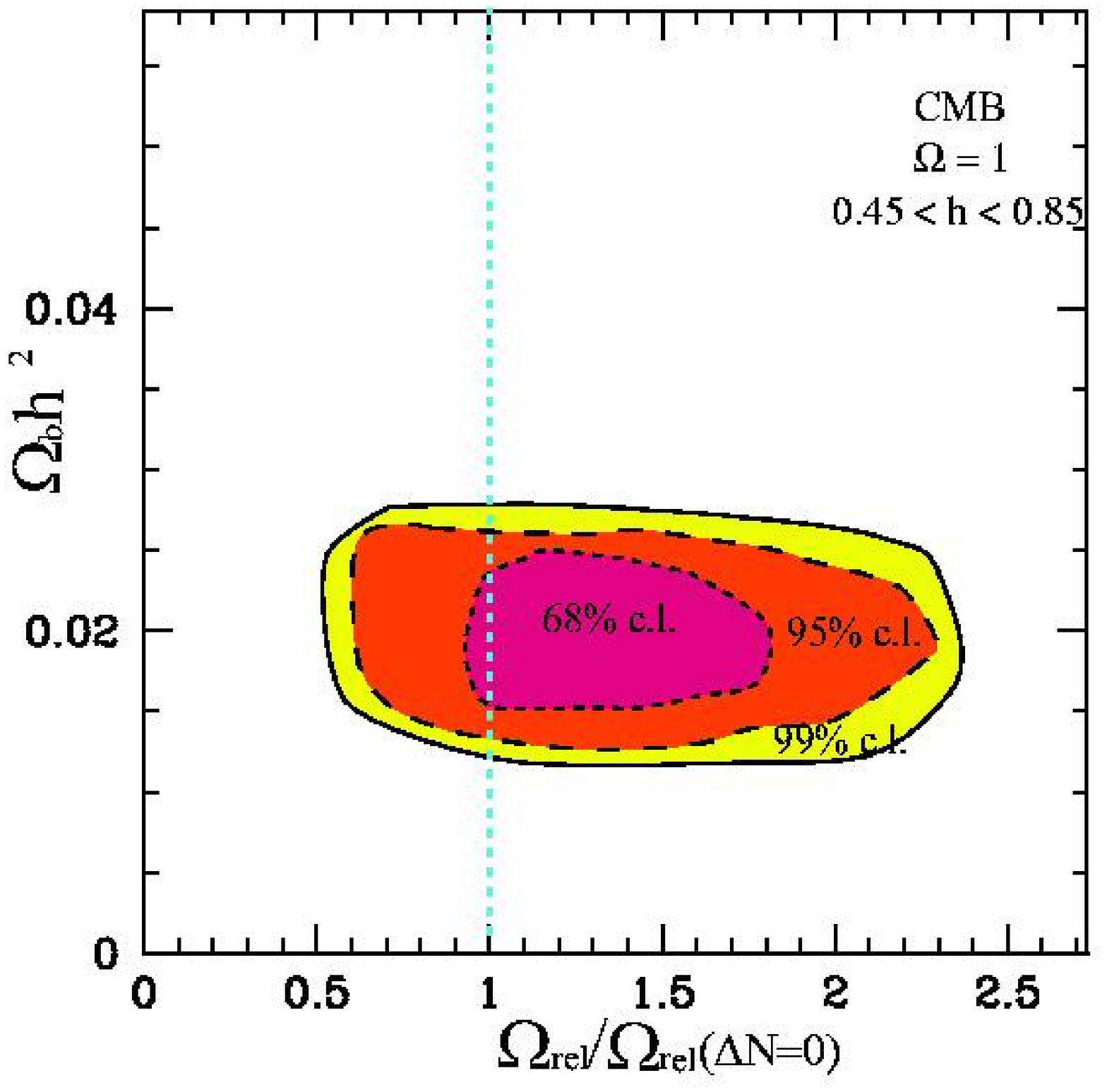}}}
{\centerline{\vbox{\epsfxsize=6.0cm\epsfbox{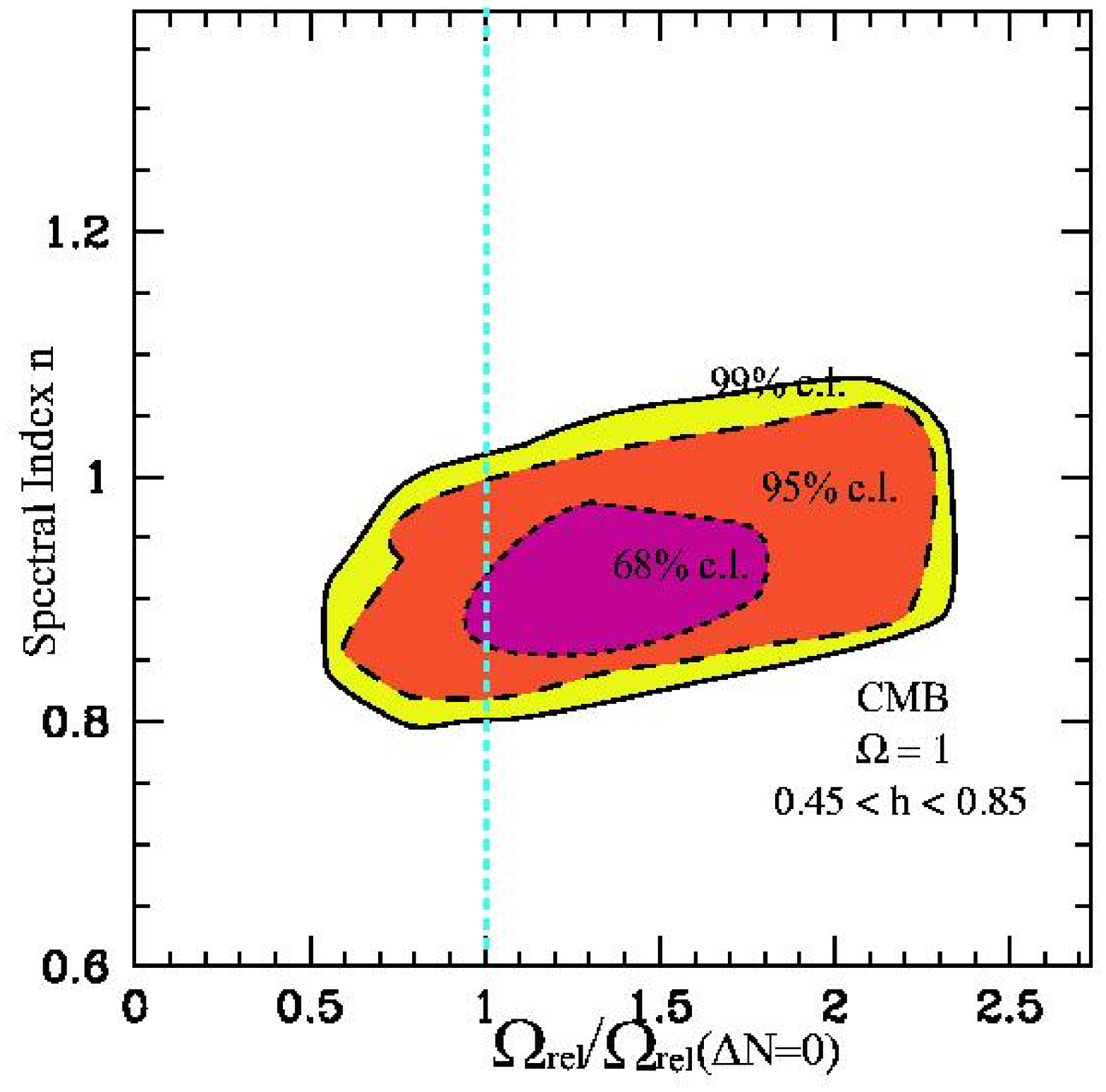}}}
{\small F{\scriptsize IG}.~3 --- 
Likelihood contours plots in the $\omega_{rel}-\omega_{m}$, 
$\omega_{rel}-\omega_b$, $\omega_{rel}-n_s$ planes.
\label{fig:plots}}}}}
\end{figure}

In Fig.\ 3 we plot the likelihood contours for $\omega_{rel}$ vs
$\omega_m, \omega_b$ and $n_s$ (top to bottom).  As we can see,
$\omega_{rel}$ is very weakly constrained to be in the range $1 \le
\omega_{rel}/\omega_{rel}(\Delta N=0) \le 1.9$ at $1-\sigma$ in all the
plots. The degeneracy between $\omega_{rel}$ and $\omega_{m}$ is
evident in the top panel of Fig.\ 3.  Increasing $\omega_{rel}$
shifts the epoch of equality and this can be compensated only by a
corresponding increase in $\omega_m$. It is interesting to note that
even if we are restricting our analysis to flat models, the degeneracy
is still there and that the bounds on $\omega_m$ are strongly
affected.  We find $\omega_m=0.2\pm0.1$, to be compared with
$\omega_m=0.13\pm0.04$ when $\Delta N$ is kept to zero.  It is important
to realize that these bounds on $\omega_{rel}$ appear because of our
prior on $h$ and because we consider flat models. When one allows $h$
as a free parameter and any value for $\Omega_m$, then the degeneracy
is almost complete and there are no bounds on $\omega_{rel}$.  In the
middle and bottom panel of Fig.\ 3 we plot the likelihood contours for
$\omega_b$ and $n_s$.  As we can see, these parameters are not
strongly affected by the inclusion of $\omega_{rel}$. The bound on
$\omega_b$, in particular, is completely unaffected by $\omega_{rel}$.
There is however, a small correlation between $\omega_{rel}$ and
$n_s$: the boost of the first peak induced by the ISW effect can be 
compensated (at least up to the third peak) by a small change in
$n_s$.

\begin{figure}
{\centerline{\vbox{\epsfxsize=8.0cm\epsfbox{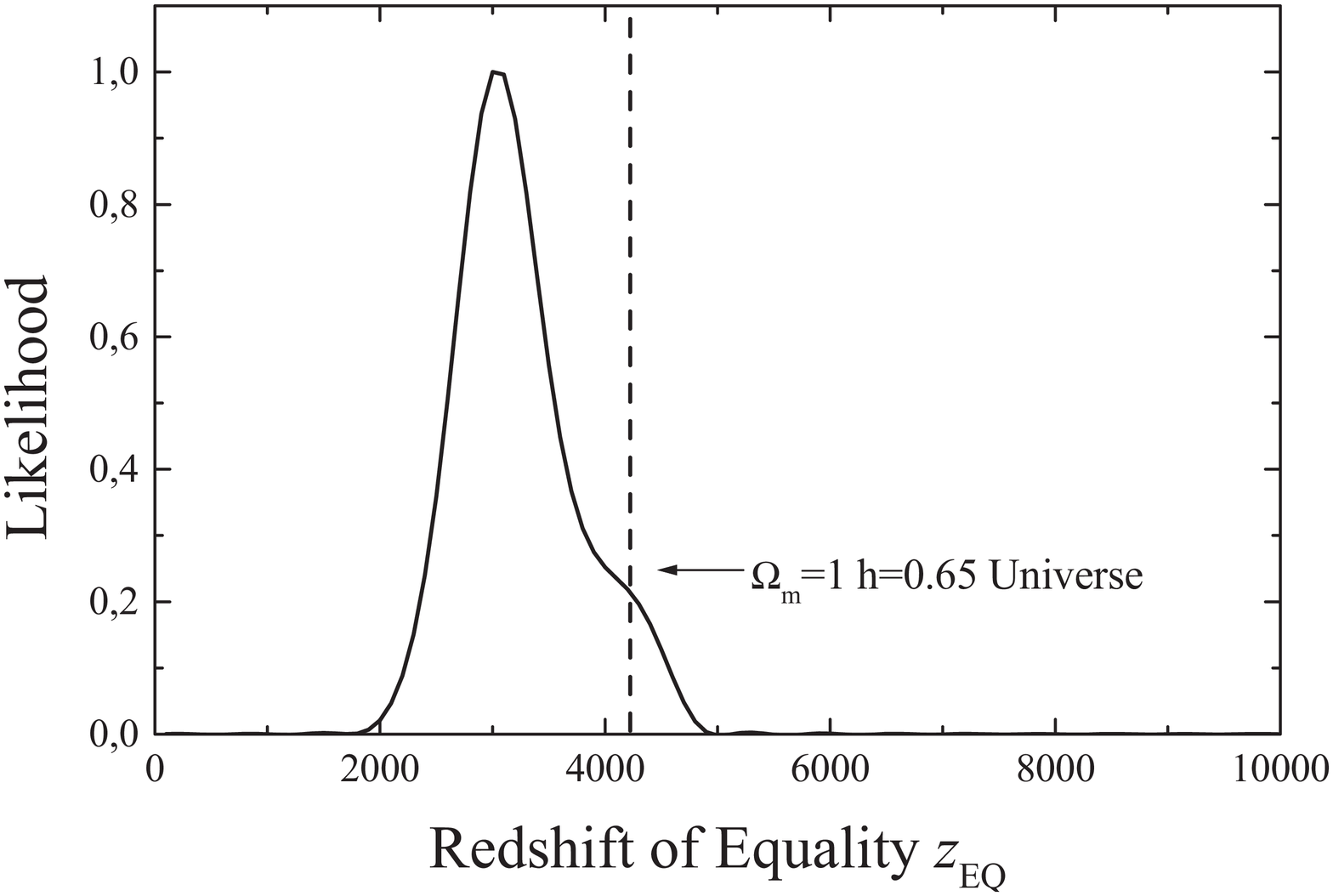}}}
{\small F{\scriptsize IG}.~4 --- Likelihood probability distribution function for
the redshift of equality.}
\label{fig:deg}}
\end{figure}

Since the degeneracy is mainly in $z_{eq}$, it is useful to estimate
the constraints we can put on this variable. In Fig.\ 4 we plot the
likelihood contours on $z_{eq}$ by using the
marginalization/maximization algorithm described above.  By
integration of this probability distribution function we obtain
$z_{eq}=3100_{-400}^{+600}$ at $68 \%$ c.l., {i.e.} a late-time
equality, in agreement with a low-density universe.

{\it External constraints.}  It is interesting to investigate how well
constraints from independent non-CMB datasets can break the above
degeneracy between $\omega_{rel}$ and $\omega_m$.  The
supernovae luminosity distance is weakly dependent on $\omega_{rel}$
(see however Zentner \& Walker 2001),
and the bounds obtained on $\Omega_m$ can be used to break the CMB
degeneracy.  Including the SN-Ia constraints on the
$\Omega_m-\Omega_{\Lambda}$ plane,
$0.8\Omega_m-0.6\Omega_{\Lambda}=-0.2\pm0.1$ (Perlmutter et al. 1999), we find
$\omega_{rel}/\omega_{rel}(\Delta N=0) =1.12_{-0.42}^{0.35}$ at the $2-\sigma$
confidence level.  

It is also worthwile to include constraints from galaxy clustering and 
local cluster abundances.
The shape of the matter power spectrum in the linear regime for
galaxy clustering can be characterized by the shape parameter 
$\Gamma \sim \Omega_mh/\sqrt{(1+0.135\Delta N)}e^{-(\Omega_b(1+
\sqrt{2h}/\Omega_m)-0.06)}$.
From the observed data one has roughly (see e.g., Bond \& Jaffe 1998) 
$0.15 \le \Gamma +(n_s-1)/2 \le 0.3$.

The degeneracy between $\omega_m$ and $\omega_{rel}$ in the CMB cannot 
be broken trivially by inclusion of large-scale structure data, because a 
similar degeneracy affects the LSS data as well (see e.g. Hu et al 1999).
However, the geometrical degeneracy is lifted in the matter power
spectrum, and accurate measurements of galaxy 
clustering at very large scales can distinguish between various models.
This is exemplified in the bottom panel of
Fig.~1, where we plot 3 matter power spectra with the same
cosmological parameters as in the top panel,  togheter with the decorrelated
matter power spectrum obtained from the PSCz survey.

The inclusion of the above (conservative) value on $\Gamma$ gives
$\omega_{rel}/\omega_{rel}(\Delta N=0) = 1.40_{-0.56}^{0.49}$, that
is less restrictive than the one obtained with the SN-Ia prior.

A better constraint can be obtained by including a prior on
the variance of matter perturbations over a sphere of size $8 h^{-1}$ Mpc,
derived from cluster abundance observations.
Comparing with $\sigma_8=(0.55\pm0.05)\Omega_m^{-0.47}$, we obtain
$\omega_{rel}/\omega_{rel}(\Delta N=0) = 1.27_{-0.43}^{0.35}$, again at
the $2-\sigma$ confidence level.


\section{Forecast for Map and Planck}

In this section we perform a Fisher matrix analysis in order to
estimate the precision with which forthcoming satellite experiments
will be able to constrain the parameter $z_{eq}$.

{\it Fisher matrix.}  Using $\Like (\mathbf{s})$ to denote the
likelihood function for the parameter set $\mathbf{s}$ and expanding
$\ln \Like $ to quadratic order about the maximum defined by the
reference model parameters $\mathbf{s_0}$, one obtains
$$
\Like \approx \Like ({\mathbf s_0}) \exp ( -\frac{1}{2} \sum_{i,j}
F_{ij} \delta s_i \delta s_j )
$$
where the {\it Fisher matrix} $F_{ij}$ is given by the expression
\begin{equation}
F_{ij} = \sum_{\ell}^{\ell_{max}} \frac{1}{(\Delta C_\ell)^2}
	 \frac{\partial C_\ell}{\partial s_i}\frac{\partial C_\ell}{\partial s_j}
\label{eq:fisher}
\end{equation}
and $\ell_{max}$ is the maximum multipole number accessible to the
experiment.  The quantity $\Delta C_\ell$ is the standard deviation 
on the estimate of $C_{\ell}$, which takes into account
both cosmic variance and the expected error of the experimental
apparatus and is given by
\begin{eqnarray}
&& (\Delta C_\ell)^2 \approx \frac{2}{(2 \ell + 1) f_{sky}}
	( C_\ell + \overline{\B}_{\ell}^{-2})^2, \\
&&  \overline{\B}_{\ell}^2 = \sum_c w_c e^{(- \ell (\ell +1)/\ell_c^2)}
\end{eqnarray}
(Knox 1995; Efstathiou \& Bond 1999), for an experiment with $N$ channels
(denoted by a subscript $c$), angular resolution (FWHM) $\theta_c$,
sensitivity $\sigma_c$ per resolution element and with a sky coverage
$f_{sky}$.
 The inverse weight per solid angle is
$w_c^{-1} \equiv (\sigma_c \theta_c)^{-2}$ and $\ell_c \equiv \sqrt{8 \ln
2}/\theta_c$ is the width of the beam, assuming a Gaussian
profile. If the initial fluctuations are Gaussian and a uniform prior
is assumed, one finds that the covariance matrix is given by the
inverse of the Fisher matrix, $C = F^{-1}$ \cite{EBT}. The standard
deviation for the parameter $s_i$ (with marginalization over all other
parameters) is therefore given by $\sigma_i = \sqrt{C_{ii}}$. This
approximation is rigorously valid only in the vicinity of the maximum
of the likelihood function, but it has proved to give useful insight
even for large values of $\mathbf{s - s_0}$ (Efstathiou \& Bond 1999;
Efstathiou 2001).
The main advantage of the Fisher matrix approach when compared to an
exact likelihood analysis is that for $m$ cosmological parameters the
former requires only the evaluation of $m+1$ power spectra. Therefore
the computational effort is vanishingly small with respect to the one
necessary for a full likelihood analysis of the parameter space.

\begin{figure}
{\centerline{\vbox{\epsfxsize=6.5cm\epsfbox{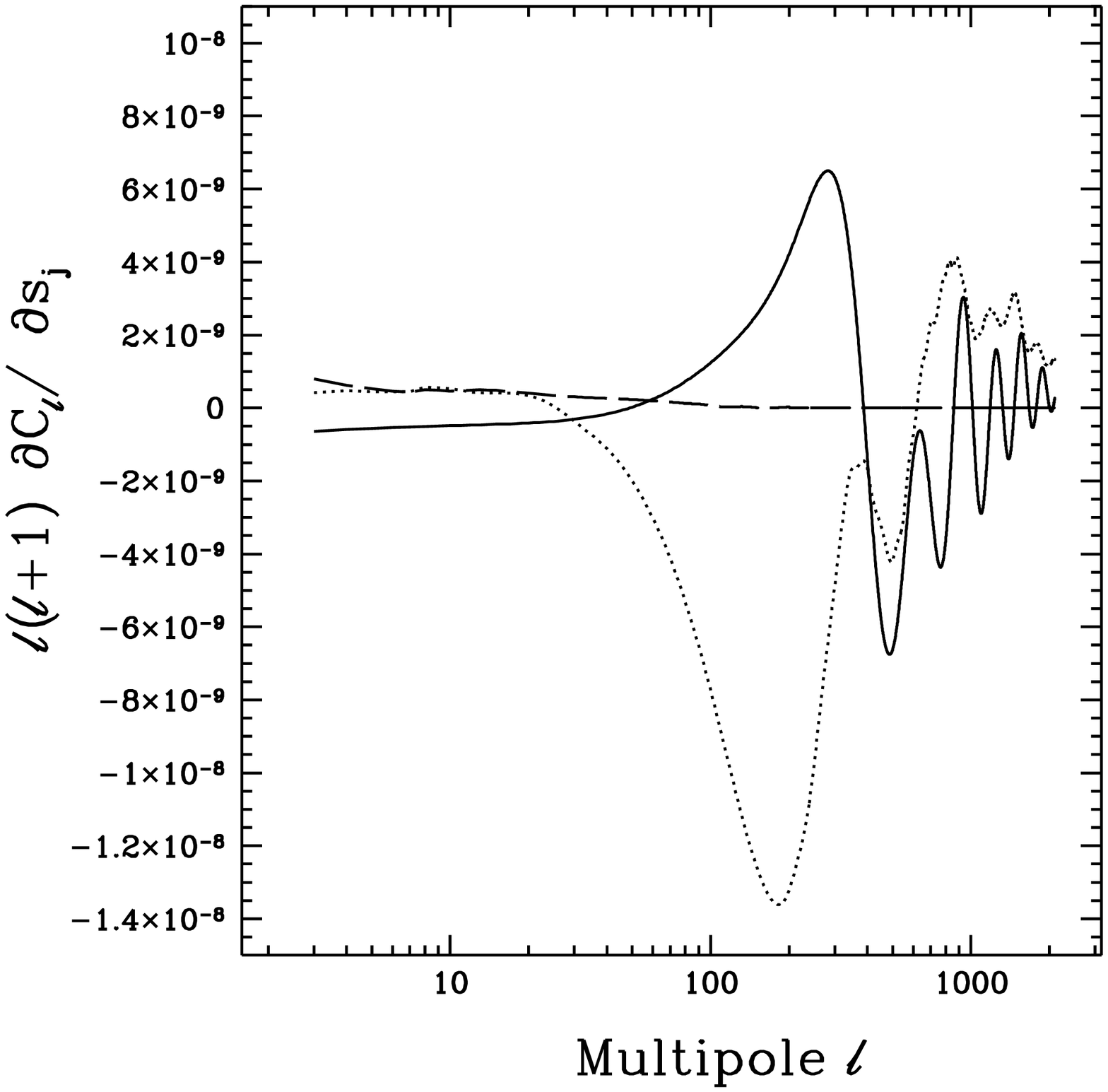}}}}
{\centerline{\vbox{\epsfxsize=6.5cm\epsfbox{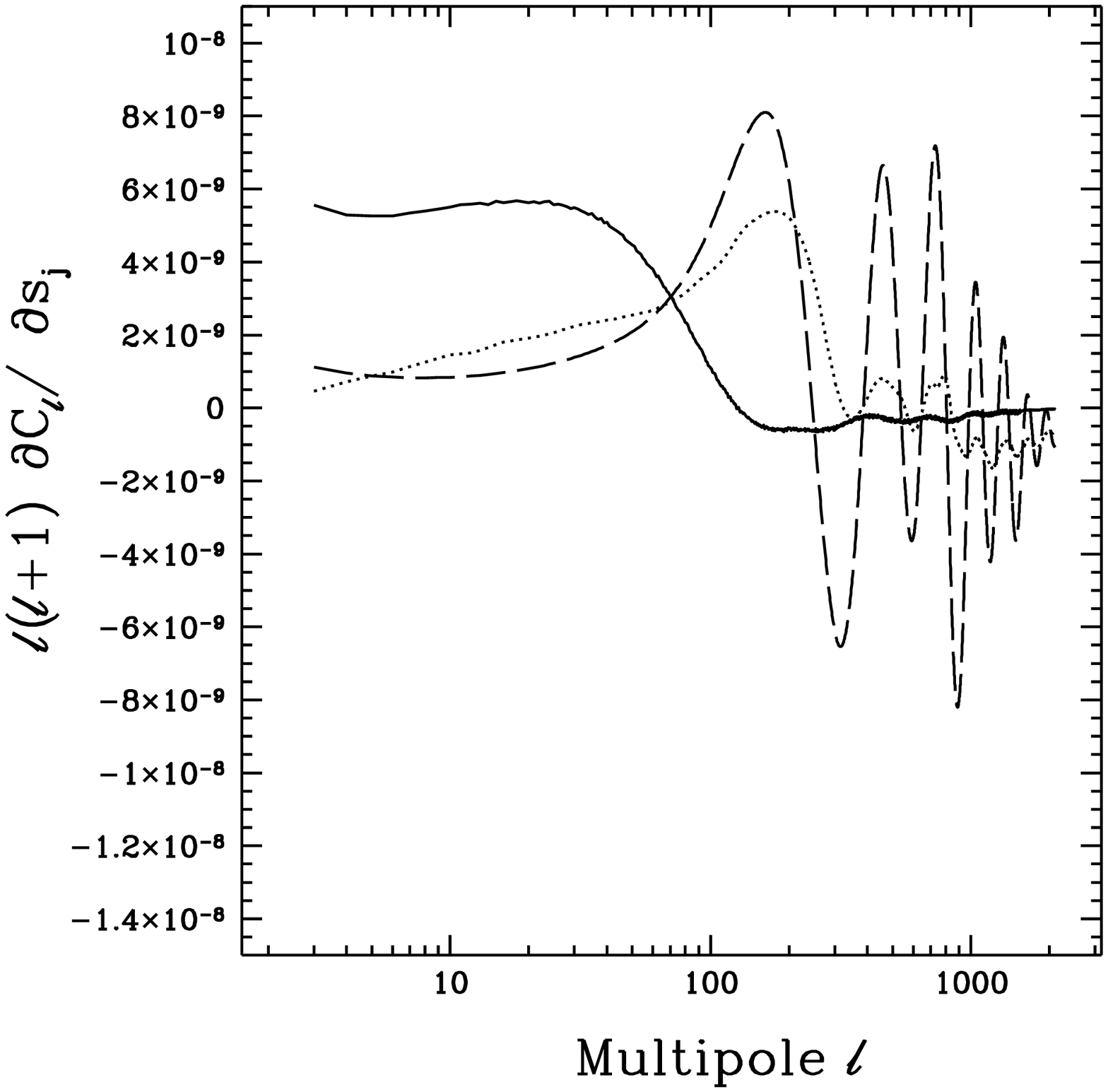}}}}
{\centerline{\vbox{\epsfxsize=6.5cm\epsfbox{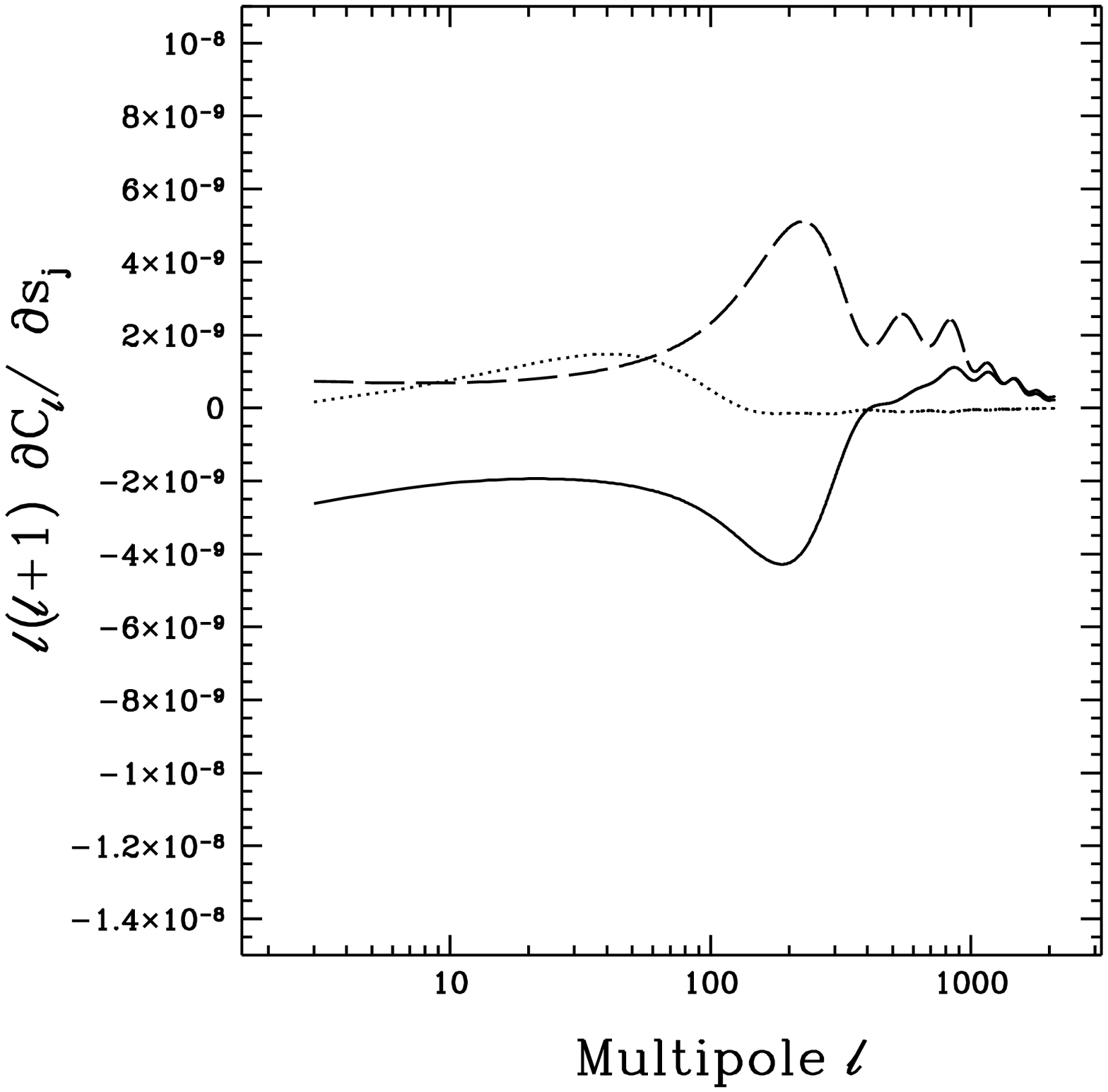}}}}
{\small F{\scriptsize IG}.~5 --- 
Derivatives of $C_\ell$ with respect to the 9 parameters
evaluated at the reference model described in the text. The derivative
$\partial C_\ell / \partial \omega_\Lambda$ has been set to 0 for
$\ell > 200$ in order to suppress the effect of numerical errors, thus
taking into account the geometrical degeneracy.  Fig.\ (a):
$0.1 \cdot \partial C_{\ell} / \partial \omega_b$ (red, solid);
$\partial C_\ell / \partial z_{eq}$ (green, dotted); $\partial C_\ell
/ \partial \omega_\Lambda$ (blue, dashed).  Fig.\ (b): $10
\cdot \partial C_\ell / \partial r $ (yellow, solid); $\partial C_\ell
/ \partial \omega_c $ (green, dotted); $\partial C_\ell / \partial \R
$ (black, dashed).  Fig.\ (c): $\partial C_\ell / \partial
n_s $ (light blue, solid); $ 10 \cdot \partial C_{\ell} / \partial n_t
$ (magenta, dotted); $ \partial C_{\ell}/\partial Q $ (black,
dashed).}
\label{fig:der}
\end{figure}

Table 1 summarizes the experimental parameters for MAP and Planck we
have used in the analysis. For both experiment we have taken
$f_{sky} = 0.50 $. These values are indicative of the expected
performance of the experimental apparatus, but the actual values may
be somewhat different, especially for the Planck satellite.

\begin{table}
\begin{tabular}{|l|ccc|cccc|}
\hline
& \multicolumn{3}{c|}{MAP}& \multicolumn{4}{c|}{Planck} \\\hline
$\nu$ (GHz) &  $40$  &  $60$  & $90$ & 
              $100$  &  $150$  & $220$ & $350$ \\
$\theta_c$ (degrees)&  $0.46$ & $0.35$ & $0.21$ &
		      $0.18$ & $0.13$ & $0.09$ & $0.08$ \\
$\sigma_c/10^{-6}$  & $6.6$  & $12.1$ & $25.5$ &   
		      $1.7$  & $2.0$  & $4.3$  & $14.4$ \\
$w^{-1}_c/10^{-15}$ & $2.9$  & $5.4$  & $6.8$  & 
		      $0.028$ & $0.022$ & $0.047$ & $0.44$  \\
$\ell_c$            & $289$ & $385$  & $642$  &
	              $757$ & $1012$ & $1472$ & $1619$ \\\hline
& \multicolumn{3}{c|}{$\ell_{max} = 1500$}
& \multicolumn{4}{c|}{$\ell_{max} = 2000$} \\\hline
\end{tabular}
\label{table:expe}

\vspace{\baselineskip}
{\small T{\scriptsize ABLE}~1 --- Experimental parameters used in the Fisher matrix analysis.}
\end{table}

{\it Cosmological parameters.} The validity of the Fisher matrix
analysis depends on the chosen parameter set, as well as on the point
${ \mathbf {s_0}}$ at which the likelihood function is supposed to
reach its maximum. We use the following 9 dimensional parameter set:
$\omega_b, \omega_c, \omega_\Lambda, \R, z_{eq}, n_s, n_t, r, Q$. Here
$n_s, n_t$ are the scalar and tensor spectral indices respectively, $r
= C_2^T/C_2^S$ is the tensor to scalar ratio at the quadrupole, and \mbox{$Q
= < \ell (\ell + 1) C_\ell > ^{1/2}$} denotes the overall
normalization, where the mean is taken over the multipole range
accessible to the experiment.
We choose to use the shift parameter $\R$ because this takes into
account the geometrical degeneracy between $\Omega_\Lambda$ and
$\Omega_k$ \cite{efs}.  Our purely adiabatic reference model has
parameters: $\omega_b = 0.0200$ ($\Omega_b = 0.0473$), $\omega_c =
0.1067$ ($\Omega_c = 0.2527$), $\omega_\Lambda = 0.2957$
($\Omega_\Lambda = 0.7000$), ($h = 0.65$), $\R = 0.953$, $z_{eq} =
3045$, $n_s = 1.00$, $n_t = 0.00$ , $r = 0.10$, $Q = 1.00$. This is a
fiducial, concordance model, which we believe is in good agreement
with most recent determinations of the cosmological parameters (flat
universe, scale invariant spectral index, BBN compatible baryon
content, large cosmological constant). Furthermore, we allow for a
modest, $10 \%$ tensor contribution at the quadrupole in order to be
able to include tensor modes in the Fisher matrix analysis.

We plot the derivatives of $C_\ell$ with respect to the different
parameters in Fig.\ 5. Generally, we remark that 
derivatives with respect to the combination of parameters 
describing the matter content of the universe ($\omega_b$ and $\omega_c$,
$\R$, $z_{eq}$) are large in the acoustic peaks region, $\ell >
100$, while derivatives with respect to parameters describing the 
tensor contribution ($n_t$, $r$) are important in the large
angular scale region. Since measurements in this region are cosmic
variance limited, we expect uncertainties in the latter set of
parameters to be large regardless of the details of the
experiment. The curve for $ \partial C_{\ell}/\partial Q $ is of
course identical to the $C_{\ell}$'s themselves. The cosmological
constant is a notable exception: variation in the
value of $\omega_\Lambda$ keeping all other parameters fixed produces
a perfect degeneracy in the acoustic peaks region. Therefore we expect
the derivative $\partial C_\ell / \partial \omega_\Lambda$ to be 0 in
this region. Small numerical errors in the computation of the spectra,
however, artificially spoil this degeneracy, erroneously leading to
smaller predicted uncertainties. In order to suppress this effect, we
set $\partial C_\ell / \partial \omega_\Lambda = 0$ for $\ell > 200$.
From eq. (\ref{eq:fisher}) we see that a large absolute value of
$\partial C_\ell/\partial s_i$ leads to a large $F_{ii}$ and
therefore to a smaller $1-\sigma$ error (roughly neglecting
non-diagonal contributions).  If the derivative along $s_i$ can be
approximated as a linear combination of the others, however, then the
corresponding directions in parameter space will be degenerate, and
the expected error will be important. This is the case for mild,
featureless derivatives as $\partial C_\ell/\partial r$, while
wild changing derivatives (such as $\partial C_\ell/\partial \R$)
induce smaller errors in the determination of the corresponding
parameter.  Therefore the choice of the parameter set is very
important in order to correctly predict the standard errors of the
experiment.

{\it Error forecast.} Table 2 shows the results of our analysis for
the expected $1 - \sigma$ error.  Determination of the redshift of
equality can be achieved by MAP with $23 \%$ accuracy, while Planck
will pinpoint it down to $2 \%$ or so. From $\omega_{rel} = (\omega_b
+ \omega_c)/z_{eq}$ it follows that the energy density of relativistic
particles, $\omega_{rel}$, will be determined within $43 \%$
by MAP and $3 \%$ by Planck.  This translates into an
impossibility for MAP of measuring the effective number of relativistic
species ($\Delta N_{eff} \approx 3.17$ at 1$\sigma$), while Planck will be able to track it down to
$\Delta N_{eff} \approx 0.24$.  As for the other parameters, while the
acoustic peaks' position (through the value of $\R$) and the matter
content of the universe can be determined by Planck with high accuracy
(of the order of or less than one percent), the cosmological constant
remains (with CMB data only) almost undetermined, because of the
effect of the geometrical degeneracy. The scalar spectral index $n_s$
and the overall normalization will be well constrained already by MAP
(within $15 \%$ and $1 \%$, respectively), while because of the
reasons explained above the tensor spectral index $n_t$ and the tensor
contribution $r$ will remain largely unconstrained by both
experiments. Generally, an improvement of a factor 10 is to be
expected between MAP and Planck in the determination of most
cosmological parameters.  Our analysis considers temperature
information only, while it is well known that inclusion of
polarization measurements greatly improves determination of the tensor
mode parameters (see e.g. Bucher et al. 2000). For Planck, which will
have polarization measurement capabilities, this will be of great
importance.

\begin{table}
\begin{tabular}{|c c c|}
\hline
                                         & MAP         & Planck \\\hline
$\delta \omega_b /\omega_b$              & 0.116       & 0.005  \\
$\delta \omega_c /\omega_c$              & 0.499       & 0.037  \\ 
$\delta \omega_\Lambda / \omega_\Lambda$ & 3.396       & 1.715  \\
$\delta \R $                             & 0.008       & 0.001  \\\hline
$\delta z_{eq} / z_{eq}$                 & 0.232       & 0.022  \\
$\delta \omega_{rel} / \omega_{rel}$     & 0.428       & 0.032  \\
$\Delta N_{eff}$                         & 3.170       & 0.237  \\\hline
$\delta n_s$                             & 0.149       & 0.013  \\
$\delta n_t$                             & 1.961       & 1.076  \\ 
$\delta r / r$                           & 5.222       & 2.670  \\     
$\delta Q$                               & 0.012       & 0.005  \\\hline
\end{tabular}
\label{table:err}

\vspace{\baselineskip}
{\small T{\scriptsize ABLE}~2 --- Fisher matrix analysis results: expected $1-\sigma$ errors for 
the MAP and Planck satellites. See the text for details and discussion.}
\end{table}

A Fisher matrix analysis for $\Delta N_{eff}$ was previously performed by
Lopez et al. (1999) and repeated by Kinney \& Riotto (1999) (with the
equivalent chemical potential $\xi$), and a strong degeneracy was
found between $N_{eff}, h$ and $\Omega_\Lambda$, and to lesser extent
with $\Omega_b$. We have seen here that the degeneracy really is
between $\omega_{rel}, \omega_m$ and $n$, and the degeneracy
previously observed is thus explained because they considered flat
models, where a change in $\Omega_\Lambda$ is equivalent to a change
in $\omega_m$, $\omega_m = (1 - \Omega_\Lambda -\Omega_b)h^2$. The
results of this paper, on how precisely the future satellite missions
can extract the relativistic energy density, can be translated into
approximately $\Delta N_{eff}=3.17$ ($\xi=2.4$) and $\Delta 
N_{eff}=0.24$ ($\xi=0.73$) for MAP and Planck respectively.

\section{Conclusions}

In this paper, we have examined the effect of varying the background of
relativistic particles on the cosmological parameters derived from CMB
observations.  We have found that the present constraints on the
overall curvature, $\Omega_{k}$, and tilt of primordial
fluctuations, $n_s$, are slightly affected by the inclusion of this
background. However, we have found a relevant degeneracy with the amount of
non relativistic matter $\omega_m$.  Even with relatively strong
external priors (flatness, $h=0.65\pm0.2$, age $> 10$ Gyrs) the
present CMB bound ($95 \%$ c.l.) $0.1 < \omega_m < 0.2$ spreads to $0.05
< \omega_m < 0.45$ when variations in $\omega_{rel}$ are allowed.
Specifically, without priors on $\omega_m$ (through flatness, $h$, etc)
no bounds on $N_{eff}$ can be obtained.

Another fundamental point bears on the identification of the 
best choice of parameters, i.\ e.\ parameter combinations 
which can be unambigously extracted from CMB data. 
It is of the greatest importance to realize 
which parameter set is least plagued by degeneracy problems, 
i.\ e.\ which directions in parameter space are non-flat. In the well
known case of the geometrical degeneracy, the shift parameter 
$\R$ can be determined with very high precision by measuring the position
of the peaks. The curvature and the Hubble parameters, however, are almost
flat directions in parameter space, and therefore are not ideal variables
for extraction from CMB data. In this work, we have pointed out that an 
analogous situation exists for $z_{eq}$, $\omega_{rel}$ and $\omega_m$. 
In fact, $z_{eq}$ is well determined because
it measures the physical distance to equality time, while
on the contrary $\omega_m$ is a rather ill-suited variable for CMB data, 
since it suffers from degeneracy with $\omega_{rel}$ (at least up to
the third acoustic peak). 

Fortunately, as we saw in the last section, the matter -- radiation 
degeneracy in the CMB data is present only up to the third peak 
and future space missions like Planck will be able to determine 
separately the amount of matter and radiation in the universe.

\section*{Acknowledgments}
We thank Celine Boehm, Ruth Durrer, Louise Griffiths and Alain Riazuelo 
for useful comments. RT is grateful to the University of Oxford for
the hospitality during part of this work.
RB is supported by the Fitzgerald Scholarship fund. 
SHH is supported by a Marie Curie Fellowship of the European Community
under the contract HPMFCT-2000-00607. AM is supported by PPARC. This work
is supported by the European Network CMBNET.

\label{lastpage}

\end{document}